\documentclass[aps,prl,twocolumn,showpacs,superscriptaddress,groupedaddress]{revtex4}

\usepackage{graphicx}  % needed for figures
\usepackage{dcolumn}   % needed for some tables
\usepackage{bm}        % for math
\usepackage{amssymb}   % for math
\usepackage{amsmath}
\usepackage{units}
\usepackage{times}

\renewcommand{\vec}[1]{\boldsymbol{#1}}

\begin{document}

\title{Reactive chemical doping of the $\mathbf{Bi}_2\mathbf{Se}_3$ topological insulator}

\author{Hadj M. Benia}
\email[Corresponding author; electronic address:\ ]{h.benia@fkf.mpg.de}
\affiliation{Max-Planck-Institut f\"ur
Festk\"orperforschung, 70569 Stuttgart, Germany}
\author{Chengtian Lin}
\affiliation{Max-Planck-Institut f\"ur Festk\"orperforschung,
70569 Stuttgart, Germany}
\author{Klaus Kern}
\affiliation{Max-Planck-Institut f\"ur Festk\"orperforschung,
70569 Stuttgart, Germany} \affiliation{Institut de Physique de la Mati{\`e}re Condens{\'e}e, Ecole Polytechnique
F{\'e}d{\'e}rale de Lausanne, 1015 Lausanne, Switzerland}
\author{Christian R. Ast}
\affiliation{Max-Planck-Institut f\"ur Festk\"orperforschung,
70569 Stuttgart, Germany}

\date{\today}

\begin{abstract}
Using angle resolved photoemission spectroscopy we studied the evolution of the surface electronic structure of the topological insulator $\text{Bi}_2\text{Se}_3$ as a function of water vapor exposure. We find that a surface reaction with water induces a band bending shifting the Dirac point deep into the occupied states and creating quantum well states with a strong Rashba-type splitting. The surface is thus not chemically inert, but the topological state remains protected. The band bending is traced back to Se-abstraction leaving positively charged vacancies at the surface. Due to the presence of water vapor, a similar effect takes place when $\text{Bi}_2\text{Se}_3$ crystals are left in vacuum or cleaved in air, which likely explains the aging effect observed in the $\text{Bi}_2\text{Se}_3$ band structure.
\end{abstract}

\maketitle

\begin{figure*}
\centerline{ \includegraphics[width = \textwidth]{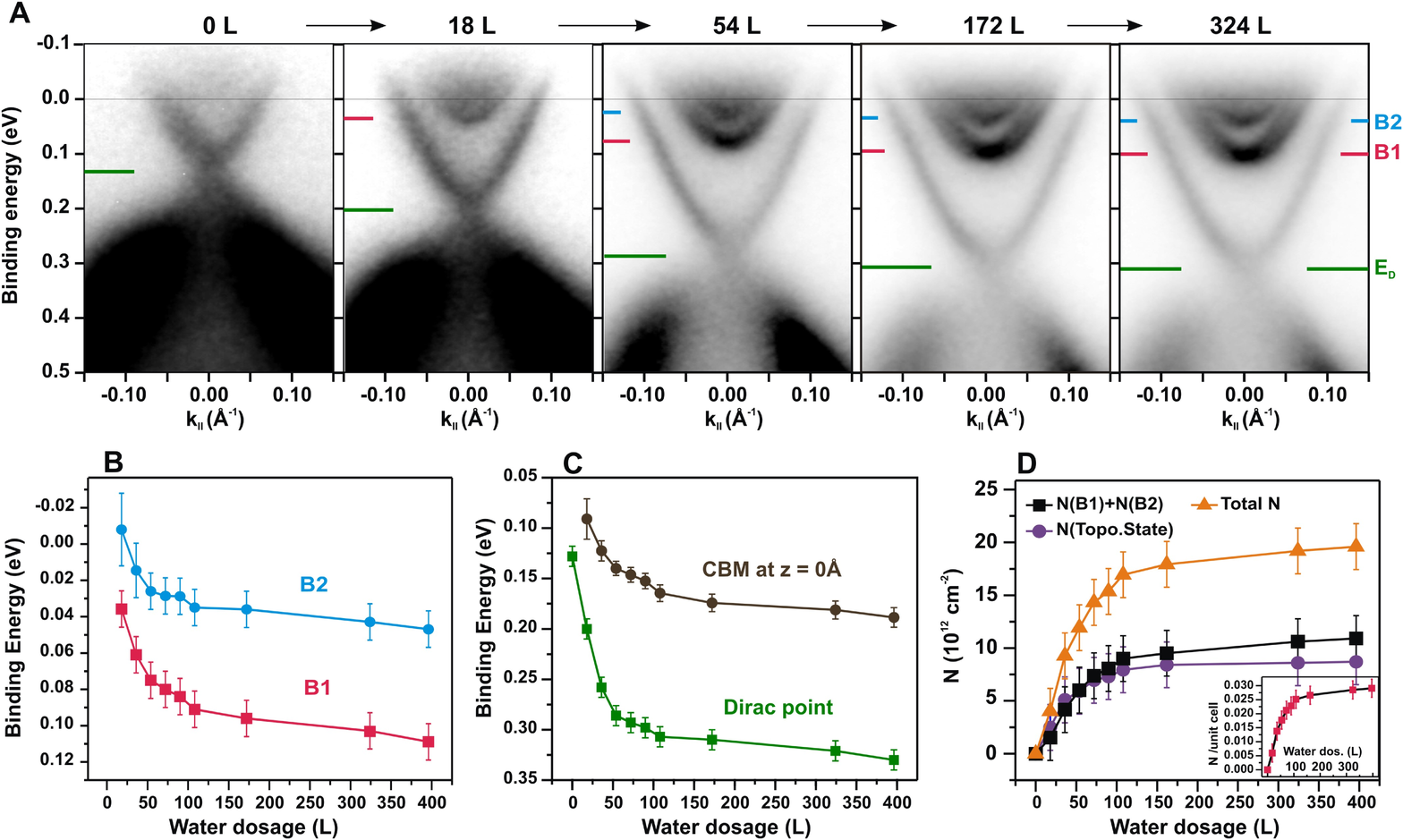}}
%\centerline{\pdfximage width \textwidth {Fig 1.png} \pdfrefximage \pdflastximage}
\caption{(a) Experimental band structure of the $\text{Bi}_2\text{Se}_3$ recorded along the $\overline{\Gamma\text{M}}$ direction showing a series of increasing exposures to water vapor (L: Langmuir). The $x$- and $y$-axes show the electron momentum $\vec{k}$ and the initial state energy $E$, respectively. The linear grayscale displays the photoemission intensity (high intensity is dark). The quantum well state bands are labelled with B1 and B2. The two branches of the topological surface state intersect at the Dirac point ($E_D$). In (b), (c), and (d) the position of the band minima (B1 and B2), the positions of the Dirac point and the conduction band minimum at the surface, and the charge carrier densities are shown as a function of water exposure, respectively. From the charge carrier densities the electron transfer per unit cell has been calculated and is shown in the inset of (d).} \label{fig:BiSeWaterExp}
\end{figure*}

Topological insulators (TIs) are a recently discovered new class of materials with a strong spin-orbit coupling which distinguish themselves from other materials through fundamental symmetry considerations \cite{fu_topological_2007,hasan_colloquium:_2010,hsieh_topological_2008}. The narrow band gap semiconductor $\text{Bi}_2\text{Se}_3$ serves as a simple model system for TIs hosting a single topologically non-trivial surface state \cite{zhang_topological_2009,hsieh_tunable_2009,xia_observation_2009}. The TIs bear great potential for the observation of exotic phenomena, such as, e.\ g., Majorana fermions, unconventional superconductivity, magnetic monopoles \cite{akhmerov_electrically_2009,fu_superconducting_2008,linder_unconventional_2010,qi_inducing_2009,tanaka_manipulation_2009}. A hallmark of the TIs is the robustness of the surface state to external perturbations. This is a consequence of the time-reversal symmetry, which requires the surface state bands to be spin degenerate at time reversal invariant points in the surface Brillouin zone. Together with the non-trivial band topology, this dictates --- on a fundamental level --- that the surface state be metallic even in the presence of small perturbations in the Hamiltonian. It has been shown already that the topologically non-trivial surface state (TSS) of $\text{Bi}_2\text{Se}_3$ is robust against different kinds of adsorbates. Depending on the adsorbate atoms or molecules, they induce an $n$-type or $p$-type doping in the surface state as a consequence of a band bending induced at the surface \cite{hor_p_2009,king_large_2011}. An important test for the suitability in device applications is the robustness of the TSS under ambient conditions. While oxygen has been shown to induce a $p$-type doping \cite{chen_massive_2010}, the effect of moisture in the air, i.\ e.\ water vapor, on the TSS in $\text{Bi}_2\text{Se}_3$ is still unknown. Water is known to react with selenides to form hydrogen selenide gas \cite{waitkins_aluminum_1946}. This point is particularly important as the surface is not just decorated with adsorbates, but in the case of water it is modified through a chemical reaction. Here, we show using angular resolved photoemission spectroscopy (ARPES) that water vapor reacts with the $\text{Bi}_2\text{Se}_3$ surface inducing an $n$-type doping of the surface. As a result, up to three quantum well states (QWS) with a strong Rashba-type splitting are formed that coexist with the robust topological state. By tuning the exposure to water vapor we can directly control the electronic structure of the $\text{Bi}_2\text{Se}_3$ surface. We also show that a similar effect related to a reaction of the surface with water molecules can be observed when $\text{Bi}_2\text{Se}_3$ crystal is exposed to residual gas in vacuum or cleaved in air. Thus, our findings can explain previously observed time-dependent effects in the surface electronic structure of $\text{Bi}_2\text{Se}_3$ \cite{hsieh_tunable_2009,bianchi_coexistence_2010,king_large_2011,fiete_how_2011}.

The $\text{Bi}_2\text{Se}_3$ crystal used in the experiments was grown following the Bridgman method using pure Bi (99.999\%) and Se (99.9999\%) materials. The ARPES measurements were done with a hemispherical SPECS HSA3500 electron analyzer characterized by an energy resolution of about 10\,meV. Monochromatized HeI (21.2\,eV) radiation was used as a photon source. During the measurements the vacuum pressure was less than $3\times 10^{-10}$\,mbar. The crystal was cleaved in vacuum at $2\times 10^{-7}$\,mbar or in air. When cleaving in air, the crystal was immediately put into a load-lock. The crystal was left \textit{as is} with no further surface cleaning procedure.

In Fig.\ \ref{fig:BiSeWaterExp}a a series of experimental band structures of the $\text{Bi}_2\text{Se}_3$ surface with increasing exposure to water vapor is shown. Measurements as well as exposure to water have been done at 300\,K. The left panel of Fig.\ \ref{fig:BiSeWaterExp}a shows the pristine $\text{Bi}_2\text{Se}_3$ surface with its TSS shortly after cleaving in high vacuum. At this point the Dirac point is at a binding energy of about 130\,meV and no conduction band is visible in the occupied states indicating a low intrinsic bulk $n$-doping. Already after a deposition of 18\,L (Langmuir) of water vapor an $n$-type doping of the TSS accompanied by the formation of an additional band B1 can be observed. Further deposition of up to 396\,L of water vapor results in the formation of a second band B2 and reveals a distinct signature of a Rashba-type spin-splitting in the first band. In all instances, the TSS is well defined and no sign of degradation is visible. We attribute the change in the electronic structure to a band bending induced by a surface modification from the water molecules. As a result, the TSS is shifted to higher binding energies and the band bending induces the formation of two-dimensional QWS, i.\ e.\ B1 and B2 at the surface. As the exposure increases the band bending increases, which further occupies the QWS and enhances their Rashba-type spin-splitting.

A more quantitative analysis of the band position for B1 and B2 is displayed in Fig.\ \ref{fig:BiSeWaterExp}b as a function of water dosage. Even though the rate at which the band shifts slows down with the exposure, a saturation for the highest exposure has not been reached at 300\,K. The same energy shift that we observe in the QWS can be observed in the shift of the Dirac point as well. In Fig.\ \ref{fig:BiSeWaterExp}c the position of the Dirac point is plotted as a function of the exposure to water vapor together with the position of the conduction band minimum (CBM) at the surface ($z=0$\,\AA). The CBM at the surface can be derived from a simple triangular potential that models the band bending \cite{Epaps}. The parallel shift of the Dirac point and the CBM indicates that the $n$-type doping of the TSS directly follows the band bending at the surface. In Fig.\ \ref{fig:BiSeWaterExp}d the change in charge carrier densities of the QWSs, the TSS as well as the sum of both are plotted as a function of water exposure. The initial charge carrier density for the TSS is 2.8$\times 10^{12}\,$cm$^{-2}$. A continuous increase can be seen as the bands continue to be occupied. From the charge carrier density we can estimate the number of electrons per unit cell that occupy the states at the surface (see inset in Fig.\ \ref{fig:BiSeWaterExp}d).

\begin{figure}
\centerline{ \includegraphics[width = 0.9\columnwidth]{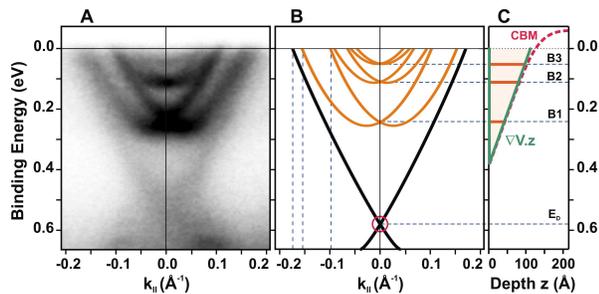}}
\caption{(A) The experimental band structure of the $\text{Bi}_2\text{Se}_3$ recorded at 100\,K along the $\overline{\Gamma\text{K}}$ direction after additional exposure to water vapor at low temperature (100\,K) for a total of 1140\,L is displayed. A third quantum well state near the Fermi level is visible. Even at this high exposure the band bending process has not saturated yet. (B) A schematic of the band structure in (A) showing the topological surface state along with the Rashba-split quantum well states. (C) Schematic of the triangular potential (solid line) extracted from the data in (A) along with the band positions (B1--B3). The band bending of the conduction band is sketched as a dashed line. The $z$-direction is perpendicular to the surface.} \label{fig:BiSeMaxWaterExp}
\end{figure}

In addition to the band bending induced by water vapor, we observe a reversible temperature-dependent change in the electronic structure, where the Dirac point and the CBM at the surface do not shift in parallel \cite{Epaps}. In order to separate these effects from each other, we have performed the measurements and the exposure to water at 300\,K. Aside from a sharpening of the bands due to a reduced electron-phonon interaction, the temperature dependence is most likely due to a combination of effects including a change of the lattice parameters. Nevertheless, cooling the sample during exposure to water further enhances the band bending effect. The experimental band structure after an additional exposure to water vapor at 100\,K for a total of 1140\,L is shown in Fig.\ \ref{fig:BiSeMaxWaterExp}a. Even after this high exposure a clear band structure of both the TSS and the QWS bands is visible. An additional band shift accompanied by an extremely strong Rashba-type spin-splitting can be observed. For this exposure a Rashba constant of  1.1\,eV\AA\ has been obtained for B1 and a third band B3 clearly appears in the QWSs. A schematic of the observed band structure is shown in Fig.\ \ref{fig:BiSeMaxWaterExp}b. The three QWS bands clearly show the Rashba-type spin-splitting with an even number of Fermi level crossings between time reversal invariant points identifying them as topologically trivial \cite{hasan_colloquium:_2010}.

Although the QWS bands and the TSS coexist at the same surface, they are of different origin. While the surface state emerges as a consequence of additional solutions to the Hamiltonian due to the presence of a surface, the QWS are confined in a potential well due to the band bending at the surface. Such QWS are not bound to the topological classification that the bulk imposes on the surface state band structure. In this regard, the QWS and the TSS should be regarded as independent with the QWS being topologically trivial.

The $\text{Bi}_2\text{Se}_3$ surface band structure shows little degradation even after exposure to several hundred Langmuir of water. This indicates a low sticking coefficient for the water vapor, but it also suggests that a reaction takes place at the surface. A likely reaction to occur is the dissociation of water followed by the formation of H$_2$Se in analogy to the reaction of Al$_2$Se$_3$ with H$_2$O \cite{waitkins_aluminum_1946}:
\begin{equation}
\text{Bi}_2\text{Se}_3 + 6\,\text{H}_2\text{O}\ \longrightarrow\ 3\,\text{H}_2\text{Se}\uparrow +\ 2\,\text{Bi}\!\left(\text{OH}\right)_3
\end{equation}
The H$_2$Se gas escapes leaving a positively charged vacancy at the surface. Selenium vacancies have been observed at the surface of $\text{Bi}_2\text{Se}_3$ in scanning tunneling microscopy (STM) measurements in ultra high vacuum (UHV) \cite{hor_p_2009,cheng_landau_2010,hanaguri_momentum-resolved_2010}. The bismuth hydroxide is left at the surface \cite{kong_rapid_2011}. The positively charged Se vacancies \cite{hor_p_2009} as well as the bismuth hydroxide contribute to an overall $n$-type doping and induce a band bending at the surface to compensate for the charge imbalance. When the band bending is strong enough, QWS are formed.

\begin{figure}
\centerline{ \includegraphics[width = 0.95\columnwidth]{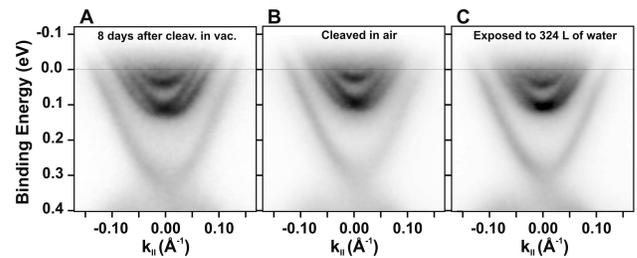}}
\caption{Experimental band structure of the $\text{Bi}_2\text{Se}_3$ after cleaving the crystal in air (A) or exposing a vacuum cleaved crystal to residual gas in a UHV chamber (B) in comparison with a controlled exposure to water vapor (C). The presence of alike Rashba-split QWS is clearly visible in all the three cases.} \label{fig:BiSeAirVacExp}
\end{figure}

A similar $n$-type doping accompanied by the formation of QWS has been observed when exposing $\text{Bi}_2\text{Se}_3$ sample to residual gas in a UHV chamber \cite{king_large_2011,bianchi_coexistence_2010}. Fig.\ \ref{fig:BiSeAirVacExp}a shows the experimental band structure of a $\text{Bi}_2\text{Se}_3$ surface exposed to the residual gas in a UHV chamber for eight days at a pressure of around $3\times10^{-10}$\,mbar. In Fig.\ \ref{fig:BiSeAirVacExp}b the band structure after being cleaved in air is shown. For comparison, the $\text{Bi}_2\text{Se}_3$ surface exposed to water vapor is also presented in Fig.\ \ref{fig:BiSeAirVacExp}c.  All three band structures clearly show the $n$-type doping of the TSS as well as two QWS with a strong Rashba-type spin-splitting. As an additional test (not shown here), we have left the cleaved $\text{Bi}_2\text{Se}_3$ surface in UHV for a total of three days at a pressure of less than $1\times10^{-11}$\,mbar. After this time no significant $n$-type doping nor an indication for a band bending have been observed. Although we cannot exclude that other adsorbates might have a similar band bending effect as water vapor, its common presence in air and in the residual gas of a UHV chamber is a strong indication that the observed effect is induced by water vapor at the $\text{Bi}_2\text{Se}_3$ surface. We therefore attribute the underlying mechanism for the band bending in all three instances of Fig.\ \ref{fig:BiSeAirVacExp} to the reaction of water vapor with the $\text{Bi}_2\text{Se}_3$ surface.

\begin{figure}
\centerline{ \includegraphics[width = 0.95\columnwidth]{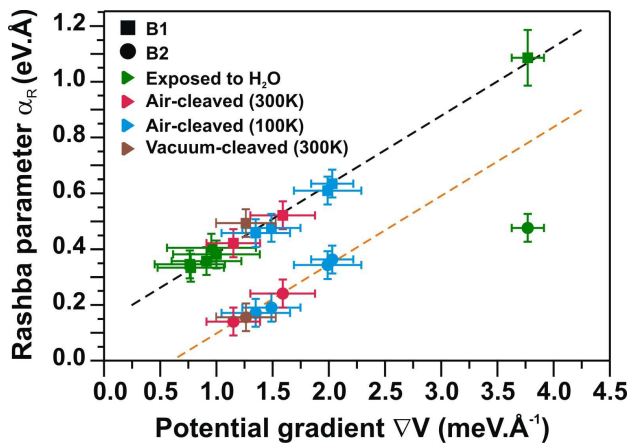}}
\caption{The Rashba constants $\alpha_R$ for different exposures to air, vacuum, as well as water vapor are plotted as a function of the effective potential gradient in the simple triangular well model. The dashed lines are a guide to the eye. The Rashba constant $\alpha_R$ correlates with the effective potential gradient regardless of how the sample was exposed.} \label{fig:AlphaRvsField}
\end{figure}

In order to analyze the Rashba-type spin-splitting of the QWS in more detail, we model the band bending at the crystal surface by a simple triangular potential with a constant effective potential gradient $\nabla V$ (\cite{Epaps}. A schematic of the potential is shown in Fig.\ \ref{fig:BiSeMaxWaterExp}c with the position of the QWS and a sketch of the conduction band bending. The strength of the Rashba-type spin-splitting can be quantified by the Rashba constant $\alpha_R=\hbar^2k_0/m^{\ast}$. The momentum offset $k_0$ as well as the effective mass $m^{\ast}$ have been directly extracted from the experimental data \cite{Epaps}. The Rashba constants $\alpha_R$ for different exposures of water are plotted for B1 and B2 in Fig.\ \ref{fig:AlphaRvsField} as a function of the effective potential gradient for different sample preparations (exposure to water vapor and air). We can see a clear correspondence between the two parameters: The larger the potential gradient, the larger the Rashba constant. Also, the Rashba constants for the different QWS bands decreases for increasing quantum number $n$, i.\ e.\ $\alpha_R(B2)<\alpha_R(B1)$. This can be understood in the way that the wave function for increasing $n$ penetrates more and more into the bulk. Furthermore, the actual potential gradient is not triangular but decreases into the bulk, although $\alpha_R(B2)<\alpha_R(B1)$ has also been shown for a triangular potential \cite{de_andrada_e_silva_spin-orbit_1997}. The potential gradient from the band bending at the surface cannot alone be responsible for the strong Rashba-type spin-splitting as has been discussed extensively in the literature \cite{petersen_simple_2000,bihlmayer_rashba-effect_2006,gierz_structural_2010}. However, it does serve as a mediator, which in combination with the strong atomic spin-orbit interaction of Bi can produce a sizeable Rashba-type spin-splitting \cite{petersen_simple_2000}. Also, the fact that the Rashba constant $\alpha_R$ behaves in a very similar way as a function of the triangular potential for the different exposures to water vapor in vacuum as well as air gives additional indirect evidence for the similar origin of the underlying mechanism.

In summary, water molecules react with the surface of $\text{Bi}_2\text{Se}_3$ crystals producing $\text{He}_2\text{Se}$ gas which leaves positively charged vacancies at the surface as well as bismuth hydroxide. This in turn induces a band bending, which eventually leads to the population of up to three QWS bands. Due to the presence of water vapor in air and in the residual gas of a UHV chamber as well as the similarity of the resulting effects, we conclude that the underlying mechanism for the band bending is a reaction of the surface with water vapor. This surface reaction with residual water is likely to be at the origin of the time-dependent band bending previously observed for $\text{Bi}_2\text{Se}_3$ crystals cleaved in UHV \cite{hsieh_tunable_2009,bianchi_coexistence_2010,king_large_2011}. The TSS undergoes an $n$-type doping and coexists with the QWS bands. It remains intact even after a chemical reaction of the surface with water or exposure to air. In principle, the surface electronic structure still qualifies as topological non-trivial as the QWS bands always add an even number of Fermi level crossings to the system and the bulk of the crystal remains insulating. However, the true quality of a TI only shows if \textit{exactly} one singly degenerate state crosses the Fermi level. In this regard, the presence of QWS states should be taken into account, for example, in transport measurements, especially for samples that have been exposed to air. Gating the sample could then allow the tuning of the Fermi level and allow a direct comparison of transport between a single and multiple band crossings in the topologically non-trivial regime.

We acknowledge stimulating discussions with M.\ Burghard, P.\ Gehring, B.\ Gao, I.\ Gierz, M.\ Assig, M.\ Etzkorn, and A.\ Schnyder.
C.\ R.\ A.\ acknowledges funding from the Emmy-Noether-Program of
the Deutsche Forschungsgemeinschaft (DFG).

\end{document}